\documentclass[11pt]{article}

\pdfoutput=1

\usepackage{comment}
\usepackage{stackrel}
\usepackage{mathrsfs}
\usepackage{fancyref}
\usepackage{graphicx}
\usepackage{enumerate}
\usepackage{mdwlist}
\usepackage{caption}
\usepackage{fancybox}
\usepackage{pifont}
\usepackage[all]{xy}
\usepackage{makecell}
\usepackage{epigraph}
\usepackage{pifont}
\usepackage{dsfont}
\usepackage{relsize}
\usepackage{enumitem}
\usepackage{float}

\usepackage{epsfig}
\usepackage{graphicx}

\usepackage{latexsym,times}
\usepackage{amssymb,amsmath}
\usepackage{amsxtra}
\usepackage{amscd}
\usepackage{amsthm}
\usepackage{amsfonts}
\usepackage{dsfont}
\usepackage{xcolor}
\usepackage{cancel}
\usepackage{framed}

\usepackage{esint}

\usepackage{enumitem}
\usepackage{float}

\usepackage{hyperref}
\usepackage[square,sort&compress,comma,numbers]{natbib}

\makeatletter

\@addtoreset{equation}{section} \makeatother

\renewenvironment{abstract}
 {\small
  \begin{center}
  \bfseries \abstractname\vspace{-.5em}\vspace{0pt}
  \end{center}
  \list{}{%
    \setlength{\leftmargin}{6mm}
    \setlength{\rightmargin}{\leftmargin}%
  }%
  \item\relax}
 {\endlist}

\newcommand{\ls}{l_{s}}

\def\PB{{{\scriptscriptstyle{\rm {P.B.}}}}}
\def\DB{{{\scriptscriptstyle{\rm {D.B.}}}}}

\newcommand{\half}{{{\textstyle\frac{1}{2}}}}

\newcommand{\be}{\begin{equation}}
\newcommand{\ee}{\end{equation} }
\newcommand{\beqa}{\begin{eqnarray} }
\newcommand{\eeqa}{\end{eqnarray} }
\newcommand{\ba}{\begin{array}}
\newcommand{\ea}{\end{array}}
\newcommand{\bpm}{\begin{pmatrix}}
\newcommand{\epm}{\end{pmatrix}}

\newcommand{\deltae}{\delta_{\epsilon}}

\newcommand{\deltaH}{\delta_{\scriptscriptstyle{H\,}}}
\newcommand{\deltaQ}{\delta_{\scriptscriptstyle{\cQ\,}}}
\newcommand{\deltaV}{\delta_{{\scriptscriptstyle{{V}}}}}

\newcommand{\rmd}{{\rm d}}

\newcommand{\rmD}{{\rm D}}

\newcommand{\ODD}{\mathbf{O}(D,D)}

\newcommand\rd{{\rm d}}
\newcommand\rD{{\rm D}}

\newcommand\cA{{\cal A}}

\newcommand\cH{{\cal H}}

\newcommand\cJ{{\cal J}}

\newcommand\cL{{\cal L}}

\newcommand\cP{{\cal P}}
\newcommand\cQ{{\cal Q}}

\newcommand\cT{{\cal T}}

\newcommand\brcP{{\bar{\cal{P}}}}

\newcommand\hcL{{\hat{\cal L}}}

\newcommand\dis{\displaystyle}

\def\tx{\tilde{x}}

\def\tpartial{\tilde{\partial}}

\newcommand{\na}{{\nabla}}

\newcommand{\bea}{\begin{eqnarray}}
\newcommand{\eea}{\end{eqnarray}}

\definecolor{rougef}{rgb}{0.56,0,0}		
\definecolor{vertf}{rgb}{0,0.5,0}		
\definecolor{bleuf}{rgb}{0,0,0.8}

\newtheorem*{theorem*}{Theorem}

\newtheorem*{lemma*}{Lemma}

\theoremstyle{definition}

\newtheorem*{definition*}{Definition}

\begin{document}

\begin{titlepage}

\title{\vskip -100pt
\vskip 2cm 
\begin{center}
A note on Faddeev--Popov action  for  doubled-yet-gauged particle\\ and  graded Poisson geometry\\
\end{center}}

\author{\sc Thomas Basile${}^{\sharp}$,\quad  Euihun Joung${}^{\sharp}$,  \quad and\quad Jeong-Hyuck Park${}^{\dagger}$}

\date{}
\maketitle 
\begin{center}
${}^{\sharp}$Department of Physics and Research Institute of Basic Science,
Kyung Hee University, \\26 Kyungheedae-ro, Dongdaemun-gu, Seoul 02447,  Korea\\\vspace{3mm}
${}^{\dagger}$Department of Physics, Sogang University, 35 Baekbeom-ro, Mapo-gu,  Seoul  04107, Korea\\
\vspace{3mm}
\texttt{ thomas.basile@khu.ac.kr\quad
euihun.joung@khu.ac.kr\quad park@sogang.ac.kr}\\
~\\

\end{center}
\begin{abstract}
\vspace{-0.2cm}  
\centering\begin{minipage}{\dimexpr\paperwidth-6.07cm}
\noindent   The section condition of Double Field Theory  has been  argued  to  mean   that    doubled coordinates are     gauged:   a gauge orbit  represents    a single physical point. In this note, we consider a    doubled and at the same time  gauged  particle action, and  show  that its   BRST formulation  including Faddeev--Popov   ghosts    matches with   the  graded Poisson geometry that has been recently used to describe the symmetries of  Double Field Theory. Besides,  by requiring  target spacetime diffeomorphisms at the quantum level, we derive quantum corrections to the classical action  involving    dilaton,   which might be comparable with  the   Fradkin--Tseytlin  term  on string worldsheet.   
\end{minipage}
\end{abstract}

\thispagestyle{empty}

\end{titlepage}


\section{Introduction}
This note  is about Double Field Theory (DFT) which was initiated with the goal  of  manifesting  the  hidden  $\ODD$ symmetry  of  supergravity~\cite{Siegel:1993xq,Siegel:1993th,Hull:2009mi,Hull:2009zb,
Hohm:2010jy,Hohm:2010pp}.  Through subsequent  further developments,  identifying the relevant     connections (Christoffel/spin) and  curvatures (scalar/Ricci/Einstein)~\cite{Jeon:2010rw,Jeon:2011cn,Jeon:2011vx,
Hohm:2011si,Park:2015bza},  it   has evolved into a stringy   `pure'   gravitational theory,    or the  $\ODD$ completion of General Relativity.   DFT  assumes the entire closed-string massless NS-NS sector  as the gravitational multiplet and   interacts  with other  sectors~\cite{Rocen:2010bk,Hohm:2011zr,Hohm:2011dv,Jeon:2012kd,
Hohm:2011ex} or   generic  matter contents~\cite{Jeon:2011kp,Choi:2015bga,Bekaert:2016isw}.  The Euler-Lagrange  equations  of the whole   NS-NS sector are unified into a single   formula, $G_{AB}=8\pi G T_{AB}$, which may well be  regarded  as the $\ODD$ completion of  the Einstein  field  equations~\cite{Angus:2018mep,Park:2019hbc}. The theory  has been shown to  admit          full order (\textit{i.e.~}quartic  in fermions)  supersymmetrizations~\cite{Jeon:2011sq,Jeon:2012hp},  and turned out to contain not only supergravity but also  various non-Riemannian gravities, \textit{e.g.~}Newton--Cartan,    as different solution sectors~\cite{Morand:2017fnv,Berman:2019izh,Blair:2019qwi,Cho:2019ofr}.

Despite the nomenclature, DFT is not truly doubled: a prescription called \textit{section condition} should be imposed on all the variables   appearing in the theory, such as   physical fields and  local parameters.   The section condition has been argued to imply that the doubled coordinates are actually gauged: a gauge orbit or an equivalence class   in the doubled coordinate space corresponds to a single physical point~\cite{Park:2013mpa}.  This idea of `coordinate gauge symmetry'  is naturally realized in sigma models where  the doubled target spacetime coordinates are  dynamical fields and thus can be genuinely  gauged~\cite{Hull:2006va,Hull:2009sg,Lee:2013hma,Bakas:2016nxt,Park:2016sbw,
Ko:2016dxa,Arvanitakis:2017hwb,Arvanitakis:2018hfn,Marotta:2018swj,
Marotta:2019wfq,Bascone:2019tuc}.  

Over the years, DFT has shown interesting and deep connections to
  various subfields of geometry, such as  Generalized
  Geometry~\cite{Hitchin:2004ut, Gualtieri:2003dx, Hitchin:2010qz,
    Cavalcanti:2011wu, Coimbra:2011nw, Coimbra:2012yy}, Courant algebroid
  (including  extensions thereof)   \cite{Chatzistavrakidis:2018ztm, Chatzistavrakidis:2019huz,
    Chatzistavrakidis:2019rpp}, and 
  para-Hermitian/Born geometry~\cite{Hull:2004in, Vaisman:2012ke,
    Vaisman:2012px, Freidel:2017yuv,
    Freidel:2018tkj, Svoboda:2018rci, Marotta:2018myj, Mori:2019slw,
    Hassler:2019wvn, Marotta:201910}. More recently, graded geometry has been 
  also  used to describe the symmetries of
  DFT~\cite{Deser:2014mxa, Deser:2016qkw, Deser:2017fko,
    Heller:2016abk, Heller:2017mwz, Stasheff:2018vnl, Deser:2018flj,
    Deser:2018oyg} (making, in particular, use of derived brackets
  introduced in~\cite{KosmannSchwarzbach1996} and further studied
  in~\cite{KosmannSchwarzbach:2003en, Voronov2004, Voronov2005, Getzler2010}).

It is the purpose of the present note to revisit the coordinate gauge symmetry
from the viewpoint of a constrained system, and along the way
 establish a  connection with the aforementioned graded geometric approach.  Specifically  we shall show that  the  doubled-yet-gauged particle 
 action constructed in \cite{Ko:2016dxa} can be formulated as a simple constrained system
 whose BRST phase space matches with  the graded manifold
 adopted  in   \cite{Deser:2018oyg,Deser:2016qkw}.  In particular,  Faddeev--Popov ghosts  carrying an $\ODD$ vector index are  mapped to    the Grassmann odd coordinates of the graded manifold. On top of that,   the   Poisson  bracket and the BRST charge    agree  with  \cite{Deser:2018oyg,Deser:2016qkw}.

The organization of the manuscript  is as follows. In the remaining of this Introduction we review the section condition,  the coordinate gauge symmetry,  and   certain elements of the graded geometric approach to DFT. Section~\ref{SECMAIN} contains our main results which split into three parts.  Firstly, we   introduce    a  constant   projector  into a section,  and  using this we reformulate the section condition as well as the coordinate gauge symmetry.    Secondly, we consider  the BRST formulation of the      doubled-yet-gauged  particle action~\cite{Ko:2016dxa}, and  point out   its   connection to  the graded  Poisson geometry~\cite{Deser:2018oyg,Deser:2016qkw}. Thirdly, by requiring  target spacetime diffeomorphisms at the quantum level, we derive quantum corrections to the classical action  involving   DFT-dilaton,  which are analogous  to  the   Fradkin--Tseytlin  term on string worldsheet.   We conclude with comments  in section~\ref{SECconclusion}. 

{\paragraph{Note added:} After the first version of our note, 
a preprint~\cite{Alfonsi:2019ggg} appeared on arXiv which discusses the coordinate gauge symmetry within the context of higher geometry.
}

\subsection*{Section condition}
DFT postulates  $\ODD$ symmetry as the first  principle  with an  invariant metric,
\be
\cJ_{AB}:={{\left(\ba{cc}{0}&{\delta^{\mu}{}_{\nu}}\\{\delta_{\rho}{}^{\sigma}}&0\ea\right)}}\,.
\label{cJ}
\ee
Along with its inverse $\cJ^{AB}$,  it can freely lower and raise  $\ODD$ vector indices (capital Roman letters).   It   decomposes     the doubled coordinates into two parts, 
\be
\ba{ll}
x^{A}=(\tx_{\mu},x^{\nu})\,,\qquad&\qquad
\partial_{A}=(\tpartial^{\mu},\partial_{\nu})\,,
\ea
\label{xtxx}
\ee
where the Greek letters are (usual) $D$-dimensional vector indices.  

In DFT, it is necessary to  impose  
  the {section condition}: 
\be
\partial_{A}\partial^{A}=0\,.
\label{SECCON}
\ee 
Acting on arbitrary functions   in the theory,  say  $\Phi_{r}$,   as well as their products like  $\Phi_{s}\Phi_{t}$, the $\ODD$ invariant D'Alembertian should vanish,  leading  to the notion of weak and strong constraints, 
\be
\ba{ll}
\partial_{A}\partial^{A}\Phi_{r}=0\quad(\mbox{weak})\,,\qquad&\qquad
\partial_{A}\Phi_{s}\partial^{A}\Phi_{t}=0\quad(\mbox{strong})\,.
\ea
\label{WSCON}
\ee
{Here we are considering  a set, $\{\Phi_{r},\Phi_{s},\Phi_{t},\cdots\}$, formed by   \textit{all} the functions in DFT  including physical fields, gauge parameters, and their derivatives (as well as numerical constants).}  The section condition is easily solved by letting
\be
\tpartial^{\mu}\equiv 0\,,
\label{particular}
\ee
such that in this case the untilde coordinates, $x^{\mu}$,  define  a section.  The general solutions to the section condition are then generated  by its     $\ODD$  rotations~\cite{Siegel:1993th,Hull:2009mi}.   Throughout  the  present manuscript, the symbol, `$\equiv$', denotes the equality   up to  $\ODD$  duality rotations,  referring to   the particular choice of the section~(\ref{particular}).

Diffeomorphisms  in  DFT consist  in  the   transformations  of the doubled coordinates,
\be
\ba{cc}
\delta x^{A}=\xi^{A}\,,
\quad&\quad
\delta\partial_{A}=-\partial_{A}\xi^{B}\partial_{B}=
(\partial^{B}\xi_{A}-\partial_{A}\xi^{B})\partial_{B}\,,
\ea
\label{diff1}
\ee
which  induce  the following transformation rule for   tensors (or tensor density  with weight $\omega$),
\be
\delta T_{A_{1}\cdots A_{n}}=-\omega\partial_{B}\xi^{B}T_{A_{1}\cdots A_{n}}+\textstyle{\sum_{i=1}^{n}\,}(\partial_{B}\xi_{A_{i}}-\partial_{A_{i}}\xi_{B})T_{A_{1}\cdots A_{i-1}}{}^{B}{}_{A_{i+1}\cdots  A_{n}}\,. 
\label{diff2}
\ee
The active version  of this  passive  transformation of  tensors  is the `generalized Lie derivative'~\cite{Siegel:1993th,Hohm:2010pp},
\be
\hcL_{\xi} T_{A_{1}\cdots A_{n}}=\xi^{B}\partial_{B}T_{A_{1}\cdots A_{n}}
+\omega\partial_{B}\xi^{B}T_{A_{1}\cdots A_{n}}+\textstyle{\sum_{i=1}^{n}\,}2\partial_{[A_{i}}\xi_{B]}T_{A_{1}\cdots A_{i-1}}{}^{B}{}_{A_{i+1}\cdots  A_{n}}\,.
\label{gLie}
\ee
Thanks to the section condition, the generalized Lie derivatives are closed  under commutator:
\be
\ba{ll}
\left[\hcL_{\zeta},\hcL_{\xi}\right]=\hcL_{\left[\zeta,\xi\right]_{\rm{C}}}\,,\quad&\qquad
\left[\zeta,\xi\right]^{M}_{\rm{C}}= \zeta^{N}\partial_{N}\xi^{M}-\xi^{N}\partial_{N}\zeta^{M}+\half \xi^{N}\partial^{M}\zeta_{N}-\half \zeta^{N}\partial^{M}\xi_{N}\,.
\ea
\label{closeda}
\ee

\subsection*{Coordinate gauge symmetry}
The section condition has been  shown to be    equivalent to   a certain translational  invariance~\cite{Park:2013mpa,Lee:2013hma},
\be
\ba{ll}
\Phi_{r}(x+\Delta)=\Phi_{r}(x)\,,\qquad&\qquad\Delta^{A}\partial_{A}=0\,,
\ea
\label{TrInv}
\ee
where   the shift parameter $\Delta^{A}$ is  `derivative-index-valued',  meaning  that its   superscript   index  should be identifiable as that of  derivative:
\be
\Delta^{A}=\Phi_{s}\partial^{A}\Phi_{t}\,.
\label{divnotion}
\ee
{Indeed, if the parameter $\Delta^A$ takes the form \eqref{divnotion}, then its contraction with a derivative vanishes by virtue of the  section condition \eqref{WSCON}. We stress that the notion of  the  `derivative-index-valuedness' is possible because  
DFT postulates  $\ODD$ symmetry and one raises (and lowers) indices using the $\ODD$-invariant metric: $\partial^{A}=\cJ^{AB}\partial_{B}$.}
The invariance of every function in DFT~(\ref{TrInv})  may suggest that the doubled coordinates are actually gauged by the shift~\cite{Park:2013mpa}:
\be
x^{A}~\sim~x^{A}+\Delta^{A}\,.
\label{EQUIV}
\ee
That is to say, each gauge orbit---or  equivalence class---corresponds to a single physical point. This idea of  `coordinate gauge symmetry'
has been applied and tested in  various  contexts. 
The finite  transformation of tensors \textit{\`{a} la} Hohm and Zwiebach~\cite{Hohm:2012gk} is equivalent to the exponentiation of the generalized Lie derivative, only  up to the equivalence relation~(\ref{EQUIV})~\cite{Park:2013mpa} (\textit{c.f.~}\cite{Berman:2014jba,Hull:2014mxa,Cederwall:2014opa}).   The usual coordinate basis of one-forms,  $\rmd x^{A}$, is not DFT-diffeomorphism covariant, 
\be
\delta(\rmd x^{A})=\rmd(\delta x^{A})=\rmd\xi^{A}=\rmd x^{B}\partial_{B}\xi^{A}\neq\rmd x^{B}(\partial_{B}\xi^{A}-\partial^{A}\xi_{B})\,.
\ee
However,  if we literally gauge  the one-form by   introducing     a derivative-index-valued connection,   it becomes   DFT-diffeomorphism covariant,
\be
\ba{llll}
\!\rmD x^{A}:=\rmd x^{A}-\cA^{A}\,,\quad&\qquad\!\! \cA^{A}\partial_{A}=0\,,\quad&\qquad\!\!
\delta(\rmD x^{A})=\rmD x^{B}(\partial_{B}\xi^{A}-\partial^{A}\xi_{B})\,,\quad&\qquad\!\!
\delta \cA^{A}=\rmD x^{B}\partial^{A}\xi_{B}\,.
\ea
\label{rmDxcA}
\ee
Further,   it is invariant  under  the coordinate gauge symmetry and thus qualifies as  a physical quantity,
\be
\ba{lll}
\delta x^{A}=\Delta^{A}\,,\qquad&\qquad\delta\cA^{A}=\rmd\Delta^{A}\,,\qquad&\qquad \delta(\rmD x^{A})=0\,.
\ea
\ee 
Using this gauged one-form, we can define a gauge invariant and (arguably) physically meaningful  
 proper length in doubled spacetime as a path integral over  the gauge connection~\cite{Park:2017snt},  recover  the doubled (and gauged) string action by Hull~\cite{Hull:2006va}\,\cite{Lee:2013hma},  and      extend  to      Green--Schwarz superstring~\cite{Park:2016sbw},  U-duality covariant  exceptional string actions~\cite{Arvanitakis:2017hwb,Arvanitakis:2018hfn} as well as    point-like    particle actions~\cite{Ko:2016dxa,Marotta:2018swj,Marotta:2019wfq,Bascone:2019tuc} (see     (\ref{Laction}) later).

\subsection*{Graded  geometric approach}
Symmetries of DFT, which are encompassed by the generalized Lie derivative~(\ref{gLie}) with  C-bracket~(\ref{closeda}), have been revisited using graded geometry in \cite{Deser:2014mxa} and further studied in \cite{Deser:2018oyg,Deser:2016qkw}. The point of this approach is to reproduce (among other things) the generalized Lie derivative, using tools from graded geometry. 
In the following, we shall introduce a few elements of this approach which are relevant to our work.
Of particular interest for us is the appearance of a graded manifold with coordinates, $\{x^{A},p_{B},\theta^{C}\}$, where $x^{A}$ and $p_{B}$ are  usual  Grassmann even variables  and $\theta^{A}$'s are odd (\textit{i.e.~}$\theta^{A}$'s anti-commute). This graded manifold is endowed with the graded Poisson bracket, 
\be
\dis{\left[F,G\right\}:=\frac{\partial F~}{\partial x^{A}}\frac{\partial G}{\partial p_{A}}-\frac{\partial F}{\partial p_{A}}\frac{\partial G ~}{\partial x^{A}}-(-1)^{{\rm deg}(F)}\frac{\partial F~}{\partial \theta^{A}}\frac{\partial G~}{\partial \theta_{A}}\,,}
\label{DB}
\ee
where ${\rm deg}(F)$ is  zero or one for even or odd  $F$ respectively.\footnote{All the derivatives  are \textit{a priori} set to act  from left,  
$
\frac{\partial F~}{\partial \theta^{A}}=\frac{\overrightarrow{\partial} \! F~}{\partial \theta^{A}}=
-(-1)^{{\rm deg}(F)}\frac{\overleftarrow{\partial}\! F~}{\partial \theta^{A}}\,,
$ 
although  for the even derivatives, $\frac{\partial\,\,\,\,}{\partial x^{A}}$ and $\frac{\partial\,\,\,\,}{\partial p_{A}}$, the ordering does not matter.  } The graded bracket is graded anti-symmetric and satisfies  the graded  Jacobi identity,  
\be
\ba{ll}
\left[F,G\right\}=-(-1)^{{\rm deg}(F){\rm deg}(G)}\left[G,F\right\}\,,\quad&\quad
\left[\left[F,G\right\},H\right\}=\left[F,\left[G,H\right\}\right\}-(-1)^{{\rm deg}(F){\rm deg}(G)}
\left[G,\left[F,H\right\}\right\}\,.
\ea
\label{Jacob}
\ee

A $p$-form (with trivial weight {$\omega=0$}) in `doubled space', $T_{A_{1}A_{2}\cdots A_{p}}=T_{[A_{1}A_{2}\cdots A_{p}]}$, can be identified   with a function in the graded manifold:
\be
T(x,\theta):=\textstyle{\frac{1}{p!}}\,T_{A_{1}A_{2}\cdots A_{p}}(x)\theta^{A_{1}}\theta^{A_{2}}\cdots\theta^{A_{p}}\,.
\label{Txtheta}
\ee
The graded  Poisson bracket then  provides\\
\indent \textit{i)\,} an inner product, 
\be
\left[\xi_{A}(x)\theta^{A},T(x,\theta)\right\}=\textstyle{\frac{1}{(p-1)!}}\,T_{BA_{1}A_{2}\cdots A_{p-1}}\xi^{B}\theta^{A_{1}}\theta^{A_{2}}\cdots\theta^{A_{p-1}}\,,
\ee
\indent \textit{ii)\,} an  expression very similar to an exterior derivative, 
\be
\,\,\,\,\,\left[p_{A}\theta^{A},T(x,\theta)\right\}=-
\textstyle{\frac{1}{p!}}\,\partial_{[A_{1}}T_{A_{2}\cdots A_{p+1}]}\theta^{A_{1}}\theta^{A_{2}}\cdots\theta^{A_{p+1}}+
\textstyle{\frac{1}{(p-1)!}}\,T_{BA_{1}A_{2}\cdots A_{p-1}}p^{B}\theta^{A_{1}}\theta^{A_{2}}\cdots\theta^{A_{p-1}}\,,
\label{almost}
\ee
\indent \textit{iii)\,} the generalized Lie derivative, 
\be
\,\,\,\,\,\left[p_{A}\theta^{A},\left[\xi_{B}\theta^{B},T\right\}\right\}+
\left[\xi_{A}\theta^{A},\left[p_{B}\theta^{B},T\right\}\right\}=
\left[
\left[p_{A}\theta^{A},\xi_{B}\theta^{B}\right\},T\right\}=
-\textstyle{\frac{1}{p!}}
\hcL_{\xi} T_{A_{1}A_{2}\cdots A_{p}}\theta^{A_{1}}\theta^{A_{2}}\cdots\theta^{A_{p}}\,,
\label{dBgLie}
\ee
\indent \textit{iv)\,} the C-bracket as a derived bracket, 
\be
\left[\left[p_{A}\theta^{A}, \xi_{B}\theta^{B} \right\}, \zeta_{C}\theta^{C}\right\} -
\left[\left[p_{A}\theta^{A}, \zeta_{B}\theta^{B} \right\}, \xi_{C}\theta^{C}\right\}
= \left[\zeta,\xi \right]_{\rm{C}}^{A}\theta_{A}\,.
\ee
The first equality  in (\ref{dBgLie})  is due to the Jacobi identity~(\ref{Jacob}),  the second holds  from  $\left[p_{A}\theta^{A},\xi_{B}\theta^{B}\right\}=p_{A}\xi^{A}-\partial_{[A}\xi_{B]}\theta^{A}\theta^{B}$, and  the resulting expression  is  analogous to the  well-known (undoubled) ``Cartan's magic formula", $\,{\mathbf{i}_{\xi}\,\rmd+\rmd\,\mathbf{i}_{\xi}=\cL_{\xi}}$.


\section{Main Results\label{SECMAIN}}
\subsection*{Formulation of the section condition  by a projector}
Here we develop some formalism which will make the  somewhat colloquial  notion, `derivative-index-valued'~(\ref{divnotion}), more concrete, and enable us to  analyze       the constrained system of the  doubled-yet-gauged sigma model powerfully later. Specifically,  we
{ describe} the section condition   by a constant projection matrix,    $\cP_{A}{}^{B}$, { along the line of 
the earlier works \cite{Hull:2004in, Hull:2006va, Hull:2009sg} and the more recent para-Hermitian approach \cite{Vaisman:2012ke, Vaisman:2012px, Freidel:2017yuv, Freidel:2018tkj, Svoboda:2018rci, Marotta:2018myj, Mori:2019slw, Hassler:2019wvn, Marotta:201910}},  as 
\be
\ba{lll}
\cP_{A}{}^{B}\cP_{B}{}^{C}=\cP_{A}{}^{C}\,,\qquad&\qquad
\cP_{AB}+\cP_{BA}=\cJ_{AB}\,,\qquad&\qquad \cP^{A}{}_{B}\partial_{A}=0\,.
\ea
\label{cPdef}
\ee
These relations  imply
\be
\cP_{A}{}^{B}\partial_{B}=\partial_{A}\,,
\label{cPpp}
\ee
and the section condition is now fulfilled  as\footnote{Alternatively from (\ref{cPdef}),  (\ref{cPbrcP}),
$\,\partial^{A}\partial_{A}=\delta_{A}{}^{B}\partial^{A}\partial_{B}=
\left(\cP_{A}{}^{B}+\brcP_{A}{}^{B}\right)\partial^{A}\partial_{B}=
\left(\cP_{A}{}^{B}\partial^{A}\right)\partial_{B}+\partial^{A}\left(\brcP_{A}{}^{B}\partial_{B}\right)=0+0=0\,.$}
\be
\partial^{A}\partial_{A}=\partial^{A}\left(\cP_{A}{}^{B}\partial_{B}\right)=
\left(\cP^{AB}\partial_{A}\right)\partial_{B}=0\,.
\ee
Its orthogonal complementary projection matrix follows 
\be
\ba{llll}
\brcP_{A}{}^{B}:=\delta_{A}{}^{B}-\cP_{A}{}^{B}=
\cP^{B}{}_{A}\,,\qquad&\qquad
\brcP_{A}{}^{B}\brcP_{B}{}^{C}=\brcP_{A}{}^{C}\,,\qquad&\qquad
\cP_{A}{}^{B}\brcP_{B}{}^{C}=0\,,\qquad&\qquad
\brcP_{A}{}^{B}\partial_{B}=0\,.
\ea
\label{cPbrcP}
\ee
The middle relation in (\ref{cPdef}) implies that the rank of the projection is $D$ as $\cJ^{AB}\cP_{AB}=\cP_{A}{}^{A}=\cP^{A}{}_{A}=D$. In other words, the section is $D$-dimensional.    Specifically,  $\brcP^{A}{}_{B}$ projects the doubled coordinates into a section $\brcP^{A}{}_{B}x^{B}$, as it satisfies the section condition  of the form~(\ref{cPdef}):  with $\partial^{A}x^{B}=\cJ^{AB}$, 
\be
\brcP_{A}{}^{B}\partial_{B}\left(\brcP^{C}{}_{D}x^{D}\right)=
\brcP_{A}{}^{B}\partial_{B}\left(\cP_{D}{}^{C}x^{D}\right)=
\brcP_{A}{}^{B}\cP_{B}{}^{C}=0\,.
\ee
Accordingly, all the  variables in DFT are functions of $\brcP^{A}{}_{B}x^{B}$ only,  independent of $\cP^{A}{}_{B}x^{B}$,   fulfilling   the section condition, ${\brcP_{A}{}^{B}\partial_{B}=0}$.  

Crucially, any derivative-index-valued vector is $\cP$-projected,  from  (\ref{cPpp}),
\be
\Phi\partial^{A}\Psi=\cP^{A}{}_{B}\left(\Phi\partial^{B}\Psi\right)\,.
\ee
Conversely, any  $\cP$-projected vector is derivative-index-valued,
\be
\cP^{A}{}_{B}V^{B}=\left(\cP_{BC}V^{C}\right)\partial^{A}\left(\brcP^{B}{}_{D}x^{D}\right)\,.
\ee
That is to say, being  $\cP$-projected  is equivalent to being  derivative-index-valued.   From now on,  we shall  always make use of  the projector  whenever it is necessary to consider the notion of  derivative-index-valuedness.  First of all, we reformulate    the coordinate gauge symmetry~(\ref{EQUIV}) and the translational invariance~(\ref{TrInv})  equivalently  as\begin{eqnarray}
&x^{A}~\sim~x^{A}+\cP^{A}{}_{B}V^{B}\,,&\label{EQUIV2}\\
&\Phi(x+\cP{ V})=\Phi(x)\quad\Longleftrightarrow\quad\cP^{A}{}_{B}\partial_{A}=0\,.&\label{trinvP}
\end{eqnarray}
Once again,   $\cP^{A}{}_{B}$ is the constant projector  of (\ref{cPdef}) and $V^{B}$ is an arbitrary variable carrying an $\ODD$ index, {\textit{c.f.} \cite{Hull:2009sg, Marotta:201910}.} \\

\noindent Explicitly for the choice of the section as $\tpartial^{\mu}\equiv0$~(\ref{particular}), or  up to $\ODD$ duality  rotations, we have\footnote{Although  constant, the skew-symmetrization of the projection matrix  may be identified  with the symplectic structure in  para-Hermitian or Born geometries~\cite{Hull:2004in, Vaisman:2012ke,
    Vaisman:2012px, Freidel:2017yuv,
    Freidel:2018tkj, Svoboda:2018rci, Marotta:2018myj, Mori:2019slw,
    Hassler:2019wvn, Marotta:201910}, as  $\,\omega_{AB}:=2\cP_{[AB]}=\cP_{AB}-\brcP_{AB}$, $\,\omega_{A}{}^{B}\omega_{B}{}^{C}=\delta_{A}{}^{C}$.} 

\be
\ba{llll}
\cP_{A}{}^{B}\equiv
\left(\ba{cc}{0}&{0}\\{0}&~\delta_{\mu}{}^{\nu}\ea\right)\,,
\quad&~~~
\cP^{A}{}_{B}\equiv\left(\ba{cc}~\delta_{\mu}{}^{\nu}&{0}\\{0}&0\ea\right)\,,
\quad&~~~
\brcP_{A}{}^{B}\equiv\left(\ba{cc}~\delta^{\mu}{}_{\nu}&{0}\\{0}&0\ea\right)\,,
\quad&~~~
\brcP^{A}{}_{B}\equiv
\left(\ba{cc}{0}&{0}\\{0}&~\delta^{\mu}{}_{\nu}\ea\right)\,,
\ea
\label{cPequiv}
\ee
such that, with (\ref{xtxx}),    $\,\cP^{AB}\partial_{B}\equiv(\partial_{\mu},0)\,$, $\,\brcP^{AB}\partial_{B}=\cP^{BA}\partial_{B}\equiv(0,\tpartial^{\mu})\equiv(0,0)\,$, and  only the tilde coordinates are gauged,
\be
 \left(\tx_{\mu}\,,\,x^{\nu}\right)~\sim~\left(\tx_{\mu}+V_{\mu}\,,\,x^{\nu}\right)\,.
\ee

\subsection*{Doubled-yet-gauged   particle action}
Now we focus on the doubled-yet-gauged  particle action constructed in \cite{Ko:2016dxa},
\be\dis{
S=\frac{1}{\,\ls}
\int\rd\tau~\half e^{-1\,}\rD_{\tau}x^{A}\rD_{\tau}x^{B}\cH_{AB}(x)-\half (m\,\ls)^{2}e\,.}
\label{Laction}
\ee
Here  $m$ is the  particle mass, $l_{s}$ is a fundamental length scale, $e$ is the  einbein,  and  
\be
\rD_{\tau}x^{A}:=\dot{x}^{A}-\cP^{A}{}_{B}A^{B}\,.
\ee
Compared to (\ref{rmDxcA}),  the derivative-index-valued gauge connection is now equivalently   set to be     $\cP$-projected as  $\cA^{A}=\cP^{A}{}_{B}A^{B}$. Further,  $\cH_{AB}$ is the DFT-metric satisfying two defining properties,
\be
\ba{ll}
\cH_{AB}=\cH_{BA}\,,\qquad&\qquad\cH_{A}{}^{C}\cH_{B}{}^{D}\cJ_{CD}=\cJ_{AB}\,,
\ea
\ee
to which  the most general solutions and thus all the possible  DFT geometries  have been  classified in \cite{Morand:2017fnv}.

The action is invariant under worldline diffeomorphisms,
\be
\ba{lll}
\deltae x^{A}=\epsilon e^{-1}\dot{x}^{A}\,,\qquad&\qquad
\deltae e=\dot{\epsilon}\,,\qquad&\qquad
\deltae A^{A}=\frac{\rd~}{\rd\tau}\left(\epsilon e^{-1} A^{A}\right)\,,
\ea
\label{wd1}
\ee
and the coordinate gauge symmetry, 
\be
\ba{lll}
\deltaV x^{A}=\cP^{A}{}_{B}V^{B}\,,\quad&\qquad
\deltaV e=0\,,\quad&\qquad
\deltaV A^{A}=\dot{V}^{A}\,.
\ea
\label{CGS}
\ee
We combine these two local symmetries, with the shift of the parameter,  $V^{A}\rightarrow V^{A}-\epsilon e^{-1}A^{A}$,
\be
\ba{lll}
\delta x^{A}=\epsilon e^{-1}\rD_{\tau}{x}^{A}+\cP^{A}{}_{B}V^{B}\,,\qquad&\qquad
\delta e=\dot{\epsilon}\,,\qquad&\qquad
\delta A^{A}=\dot{V}^{A}\,.
\ea
\label{local}
\ee
Thanks to the shift,  $\delta x^{A}$ now assumes a covariant form, while  $\delta e$ and $\delta A^{A}$ are separately given by the time derivative of each   gauge parameter.

\subsection*{Hamiltonian action}
In order to obtain  more insights into the coordinate gauge symmetry from the view point of a  constrained system, \textit{e.g.~}\cite{Henneaux:1992ig},  we reformulate  the doubled-yet-gauged particle action~(\ref{Laction})  into the 
Hamiltonian form,
\be
\dis{
S_{H}=\frac{1}{\,\ls}
\int\rd\tau~p_{A}\dot{x}^{A}-A^{A}\brcP_{A}{}^{B}p_{B}- e  H(x,p)\,,}
\label{Haction}
\ee
where the Hamiltonian is given by
\be
H(x,p)=\half   p_{A}p_{B} \cH^{AB}(x)+\half  (m\,l_{s})^{2}\,.
\label{Hamiltonian}
\ee
Now, $A^{A}$ and $e$ are  Lagrange multipliers and generate   two first-class constraints, 
\be
\ba{ll}
H_{A}:=\brcP_{A}{}^{B}p_{B}=p_{B}\cP^{B}{}_{A}\approx 0\,,\qquad&\qquad
H\approx 0\,,
\ea
\label{first}
\ee
{which Poisson-commute,
$[H_A\,H\}_{\PB}=0$, upon imposing the section condition, $\cP^{C}{}_{D}\, \partial_{C}\, \cH_{AB}=0$.}
The dynamics is governed by the total Hamiltonian, $H_{{\rm total}}=A^{A}H_{A}+ eH$, 
\be
\ba{ll}
\dot{x}^{A}=\left[x^{A},H_{{\rm total}}\right\}_{\PB}= e\cH^{AB}p_{B}+\cP^{A}{}_{B}A^{B}\,,\qquad&\quad
\dot{p}_{A}=\left[p_{A},H_{{\rm total}}\right\}_{\PB}=-\half e p_{B}p_{C\,}\partial_{A}\cH^{BC}\,.
\ea
\ee
Integrating out the auxiliary momenta, $p_{A}$, in  (\ref{Haction})  one  recovers   (\ref{Laction}).   Surely, the two first-class constraints reflect the underlying two gauge symmetries of the Hamiltonian action~(\ref{Haction}),   the coordinate gauge symmetry and the worldline diffeomorphisms:\footnote{Recall  that any Hamiltonian action, $S[x^{a},p_{b},\lambda^{i}]=\dis{\int}\rd t~p_{a}\dot{x}^{a}-\lambda^{i}\chi_{i}(x,p)$, with first-class constraints, $\chi_{i}\approx 0$  obeying $\{\chi_{i},\chi_{j}\}_{\PB}=f^{k}{}_{ij}\chi_{k}$, has the gauge symmetry,
\[
\ba{lll}
\delta x^{a}=\{x^{a},\chi_{i}\}_{\PB}\epsilon^{i}\,,\qquad&\quad
\delta p_{a}=\{p_{a},\chi_{i}\}_{\PB}\epsilon^{i}\,,\qquad&\quad
\delta \lambda^{i}=\dot{\epsilon}^{i}-f^{i}{}_{jk}\lambda^{j}\epsilon^{k}\,.
\ea
\]
In our case, $f^{i}{}_{jk}=0$ as it is  Abelian~(\ref{Abelian}).  } 
\be
\ba{llll}
\deltaH x^{A}=\epsilon\cH^{AB}p_{B}+\cP^{A}{}_{B}V^{B}\,,\qquad&\quad
\deltaH p_{A}=-\half\epsilon\partial_{A}\cH^{BC}p_{B}p_{C}\,,\qquad&\quad
\deltaH e=\dot{\epsilon}\,,\qquad&\quad
\deltaH A^{A}=\dot{V}^{A}\,.
\ea
\label{deltaH}
\ee
The difference between (\ref{local}) and (\ref{deltaH}) amounts to the so-called trivial gauge symmetry~\cite{Henneaux:1992ig}.   It is important to remark that the former  constraint in (\ref{first})  is projected,
\be
\brcP_{A}{}^{B}H_{B}=H_{B}\cP^{B}{}_{A}=H_{A}\,.
\ee

\subsection*{BRST formulation}
Finally, let us extend the classical action, $S$ in (\ref{Laction}),  to the  
Faddeev--Popov gauge-fixed   action,   
\be\dis{
S_{{\scriptscriptstyle{\rm{F.P.}}}}=\frac{1}{\,\ls}
\int\rd\tau~\half e^{-1\,}\rD_{\tau}x^{A}\rD_{\tau}x^{B}\cH_{AB}(x)-\half (m\ls)^{2}e
+k_{A}\cP^{A}{}_{B}A^{B}+k(e-1)+\half \theta_{A}\dot{\theta}^{A}+\sum_{\alpha=1}^{2}
\half\vartheta_{\alpha}\dot{\vartheta}^{\alpha}\,,}
\label{BRSTaction}
\ee
where $\theta^{A}$ and $\vartheta^{\alpha}$ with $\alpha=1,2$ are Grassmann  odd variables. 
Readers might not immediately recognize  standard  ghost terms, but the above action contains them precisely and  we have a (good) reason to spell the  action  like above,  which we explain shortly. 

First of all, integrating out the auxiliary variables, $k_{A}$, $k$, we are fixing the gauge,
\be
\ba{ll}
\cP^{A}{}_{B}A^{B}=0\,,\qquad&\qquad e=1\,.
\ea
\label{gaugefix}
\ee
Decomposing the odd variable, $\theta^{A}$, into two parts, 
\be
\ba{lll}
\theta^{A}=C^{A}+B^{A}\,,\qquad&\quad
C^{A}:=\cP^{A}{}_{B}\theta^{B}\,,\qquad&\quad
B^{A}:=\brcP^{A}{}_{B}\theta^{B}\,,
\ea
\label{BCtheta}
\ee
we may identify the standard $BC$ ghost term,  from (\ref{cPbrcP}),  (\ref{local}),  up to total derivative,\footnote{If we fix the section, $\tpartial^{\mu}\equiv0$,  with (\ref{cPequiv}), we note
\[
\ba{llllll}
\cP^{A}{}_{B}A^{B}\equiv(A_{\mu}\,,\,0)\,,\quad&
\theta^{A}\equiv(C_{\mu}\,,\,B^{\nu})\,,\quad&
C^{A}\equiv(C_{\mu}\,,\,0)\,,\quad&
B^{A}\equiv(0\,,\,B^{\nu})\,,\quad& B_{A}\dot{C}^{A}\equiv B^{\mu}\dot{C}_{\mu}\,,
\quad&\quad U\equiv C_{\mu}B^{\mu}\,.
\ea
\]\label{footnotefix}}   
\be
\half \theta_{A}\dot{\theta}^{A}=\half B_{A}\dot{C}^{A}+\half {C}^{A}\dot{B}_{A}
=B_{A}\dot{C}^{A}+\textstyle{\frac{1}{2}\frac{\rd~}{\rd\tau}}\!\left(C^{A}B_{A}\right)\,.
\label{BdotC}
\ee
The ghost number $U$ is  defined as
\be
U:=\cP_{AB}\theta^{B}\theta^{A}=C^{A}B_{A}\,,
\ee
which ranges from $-D$ to $+D$.                   

Similarly for the worldline diffeomorphisms,  we identify 
\be
\ba{ll}
\vartheta_{1}=\vartheta^{2}=b\,,\qquad&\qquad
\vartheta_{2}=\vartheta^{1}=c\,,
\ea
\ee
and   the corresponding  $bc$ ghost term,  along with the ghost number, $u$,
\be
\ba{ll}
\dis{\sum_{\alpha=1}^{2}\,\half \vartheta_{\alpha}\dot{\vartheta}^{\alpha}=b\dot{c}+\half\textstyle{\frac{\rd~}{\rd\tau}}\!\left(cb\right)\,,}\qquad&\qquad u:=cb\,.
\ea
\label{bdotc}
\ee
Intriguingly, an $\mathbf{O}(1,1)$ structure  has appeared  with  the invariant metric given  by    the second Pauli matrix, $\sigma_{2}$, which might hint at the `doubling' of the worldline, \textit{c.f.~}\cite{Bergshoeff:2019sfy}. It is also  amusing  to observe that the total derivatives in (\ref{BdotC}), (\ref{bdotc})   actually  contribute to  the action~(\ref{BRSTaction}) through    the  ghost number changes     at   boundaries as   $\textstyle{\frac{1}{2 l_{s}}}\left( U+ u\right)\big|_{\tau=-\infty}^{\tau=+\infty}\,$. \vspace{2pt}

The  BRST differential, $\deltaQ$, with the nilpotency,   $\delta_{\scriptscriptstyle{\!\cQ}}^{2}=0$,   is given by
\be
\ba{lll}
\deltaQ x^{A}=c e^{-1}\rD_{\tau}{x}^{A}+C^{A}\,,\qquad&\qquad
\deltaQ e=\dot{c}\,,\qquad&\qquad
\deltaQ A^{A}=\dot{C}^{A}\,,\\
\deltaQ B_{A}=-\brcP_{A}{}^{B}k_{B}\,,\qquad&\qquad
\deltaQ C_{A}=0\,,\qquad&\qquad\deltaQ k_{A}=0\,,\\
\deltaQ b=-k\,,\qquad&\qquad
\deltaQ c=0\,,\qquad&\qquad\deltaQ k=0\,.
\ea
\label{BRST}
\ee
On-shell we have ${\dot{\theta}^{A}=0}$, ${\dot{\vartheta}^{\alpha}=0}$,   and  these make $\deltaQ e$, $\deltaQ A^{A}$ trivial  and  thus consistent with the gauge fixing~(\ref{gaugefix}). \\

\noindent Now, we proceed to the Hamiltonian formulation  of the Faddeev--Popov action~(\ref{BRSTaction}). We   denote the canonical momenta of  $x^{A}$,  $\theta^{B}$,  $\vartheta^{i}$ by   $\,p_{A}$, $\Pi_{B}$,  $\pi_{i}$ respectively, and write  the Poisson bracket,
\be
\dis{\left[F,G\right\}_{\PB}=\frac{\partial F}{\partial x^{A}}\frac{\partial G}{\partial p_{A}}-\frac{\partial F}{\partial p_{A}}\frac{\partial G}{\partial x^{A}}+(-1)^{{\rm deg}(F)}\!\left(\frac{\partial F}{\partial \theta^{A}}\frac{\partial G}{\partial \Pi_{A}}+\frac{\partial F}{\partial \Pi_{A}}\frac{\partial G}{\partial \theta^{A}}+
\frac{\partial F}{\partial \vartheta^{\alpha}}\frac{\partial G}{\partial \pi_{\alpha}}+\frac{\partial F}{\partial \pi_{\alpha}}\frac{\partial G}{\partial \vartheta^{\alpha}}\right)\,.}
\label{PB}
\ee
The dynamics  after the gauge fixing~(\ref{gaugefix})  is governed by the Hamiltonian~(\ref{Hamiltonian}),  subject to the two first-class constraints~(\ref{first})  and further two  additional    second-class constraints, 
\be
\ba{ll}
\phi_{A}:=\Pi_{A}+\half \theta_{A}\approx 0\,,\qquad&\qquad
\left[\phi_{A},\phi_{B}\right\}_{\PB}=-\cJ_{AB}\,,\\
\varphi_{\alpha}:=\pi_{\alpha}+\half \vartheta_{\alpha}\approx 0\,,\qquad&\qquad
\left[\varphi_{\alpha},\varphi_{\beta}\right\}_{\PB}=-(\sigma_{2})_{\alpha\beta}\,.
\ea
\label{second}
\ee
The  first-class constraints form an  Abelian algebra, 
\be
\ba{lll}
\left[H_{A},H_{B}\right\}_{\PB}=0\,, \quad&\qquad
\left[H_{A},H\right\}_{\PB}=0\,,\quad&\qquad
\left[H_{A},\phi_{B}\right\}_{\PB}=0\,,\\
\left[H_{A},\varphi_{B}\right\}_{\PB}=0\,,\quad&\qquad
\left[H,\phi_{A}\right\}_{\PB}=0\,,\quad&\qquad
\left[H,\varphi_{A}\right\}_{\PB}=0\,,
\ea
\label{Abelian}
\ee
and the second-class constraints originate directly  from the  Faddeev--Popov action~(\ref{BRSTaction}).  If we had assumed the conventional $BC$ ghost terms as in (\ref{BdotC}) and (\ref{bdotc}),  the identification of the  second-class constraints would have been obscure.   This justifies the precise form of our Faddeev--Popov action~(\ref{BRSTaction}).

The  relevant Dirac bracket reads 
\be
\dis{\left[F,G\right\}_{\DB}}=
\dis{\left[F,G\right\}_{\PB}
+\left[F,\phi_{A}\right\}_{\PB}\left[\phi^{A},G\right\}_{\PB}+\left[F,\varphi_{\alpha}\right\}_{\PB}\left[\varphi^{\alpha},G\right\}_{\PB}\,.}
\label{DiracB0}
\ee
which satisfies  
\be
\ba{lll}
\left[x^{A},p_{B}\right\}_{\DB}=\delta^{A}{}_{B}\,,\qquad&\quad
\left[\theta_{A},\theta_{B}\right\}_{\DB}=\cJ_{AB}\,,\qquad&\quad
\left[\vartheta_{\alpha},\vartheta_{\beta}\right\}_{\DB}=(\sigma_{2})_{\alpha\beta}\,,\\
\left[\theta_{A},\vartheta_{\alpha}\right\}_{\DB}=0\,,\qquad&\quad
\left[\phi_{A},F\right\}_{\DB}=0\,,\qquad&\quad
\left[\varphi_{\alpha},F\right\}_{\DB}=0\,,
\ea
\label{DBr}
\ee
recovering more familiar   bracket structure of the ghost system,\footnote{For the fixed  section of  $\tpartial^{\mu}\equiv0$, with footnote~\ref{footnotefix}, $\left[B^{\mu},C_{\nu}\right\}_{\DB}=\delta^{\mu}{}_{\nu}$.}
\be
\ba{lll}
\left[B_{A},B_{B}\right\}_{\DB}=0\,,
\quad&\quad
\left[C_{A},C_{B}\right\}_{\DB}=0\,,
\quad&\quad
\left[B_{A},C_{B}\right\}_{\DB}=\left[C_{B},B_{A}\right\}_{\DB}=\bar\cP_{AB}=\cP_{BA}\,,\\
\left[b,b\right\}_{\DB}=0\,,\quad&\quad
\left[c,c\right\}_{\DB}=0\,,\quad&\quad
\left[b,c\right\}_{\DB}=\left[c,b\right\}_{\DB}=1\,.
\ea
\ee
Thus, on the surface of the second-class constraints, the Poisson bracket reduces to Dirac bracket given by
\be
\dis{\left[F,G\right\}_{\DB}=\frac{\partial F~}{\partial x^{A}}\frac{\partial G}{\partial p_{A}}-\frac{\partial F}{\partial p_{A}}\frac{\partial G ~}{\partial x^{A}}-(-1)^{{\rm deg}(F)}\!\left(
\frac{\partial F~}{\partial \theta^{A}}\frac{\partial G~}{\partial \theta_{A}}
+\frac{\partial F~}{\partial \vartheta^{\alpha}}\frac{\partial G~}{\partial \vartheta_{\alpha}}\right)\,.}
\label{DiracB}
\ee
The first-class constraints~(\ref{first}) make up the BRST charge,
\be
\ba{llll}
\cQ:=Q+q\,,\qquad&\qquad Q:=C^{A}H_{A}\,,\qquad&\qquad q:= cH\,,\qquad&\qquad
\left[\cQ,\cQ\right\}_{\DB}=0\,,
\ea
\label{cQ}
\ee
where the  nilpotency  is   ensured, with $C^{A}\partial_{A}=0$ (the section condition),  by
\be
\ba{lll}
\left[Q,Q\right\}_{\DB}=0\,,
\qquad&\qquad
\left[Q,q\right\}_{\DB}=0\,,
\qquad&\qquad
\left[q,q\right\}_{\DB}=0\,.
\ea
\ee
The charge, $\cQ$, generates  the BRST symmetry~(\ref{BRST}) through the Dirac bracket, 
\be
\ba{ll}
\left[x^{A},\cQ\right\}_{\DB}=C^{A}+c \cH^{AB}p_{B}\,,\qquad&\quad
\left[p_{A},\cQ\right\}_{\DB}=-\half c \partial_{A}\cH^{BC}p_{B}p_{C}\,,\\
\left[B_{A},\cQ\right\}_{\DB}=H_{A}\,,\qquad&\quad
\left[C^{A},\cQ\right\}_{\DB}=0\,,\\
\left[b,\cQ\right\}_{\DB}= H\,,\qquad&\quad
\left[c,\cQ\right\}_{\DB}=0\,,
\ea
\label{BRST2}
\ee
and commutes with the Hamiltonian, $\left[H,\cQ\right\}_{\DB}=0$, as should be.

Crucially, restricted on the phase  space  of $\{x^{A}, p_{B}, \theta_{C}\}$ with the trivial  $bc$ ghost number, the Dirac bracket~(\ref{DiracB})  reduces  precisely   to the graded bracket \textit{\`{a} la} Deser and S\"{a}mann~(\ref{DB}) \cite{Deser:2016qkw}.\,\footnote{Notice however the $\mathbb Z$-grading used in \cite{Deser:2016qkw} is different from 
the ghost number grading of our particle action. 
Consequently, the bracket \eqref{DiracB} has ghost number zero, 
but degree $-2$ in the grading of \cite{Deser:2016qkw}.}

\subsection*{Quantum correction}
We consider quantizing the Dirac bracket~(\ref{DBr}). We set a vacuum state, $|0\rangle$,  which is annihilated by $\hat{p}_{A}$, $\hat{B}_{A}$, and $\hat{b}$.  
 Any physical state, $|\Psi\rangle=\Psi(\hat{x})|0\rangle$, having trivial ghost numbers should satisfy 
\begin{eqnarray}
&&\!\!\!\!\!\!\!\!\!\hat{H}_{A}|\Psi\rangle=\brcP_{A}{}^{B}\hat{p}_{B}
\Psi(\hat{x})|0\rangle=-i\hbar\brcP_{A}{}^{B}\partial_{B}\Psi(\hat{x})|0\rangle=0\,,\\
&&\!\!\!\!\!\!\!\!\!\hat{H}|\Psi\rangle=\half\left(\hat{p}_{A}\cH^{AB}(\hat{x})\hat{p}_{B}+(m\, l_{s})^{2}\right)\!
\Psi(\hat{x})|0\rangle=\half\big[-\hbar^{2}\partial_{A}\!\left(\cH^{AB}(\hat{x})\partial_{B}\Psi(\hat{x})\right)+(m\,l_{s})^{2}\Psi(\hat{x})\big]|0\rangle=0\,,\nonumber\\
{}\label{latter}
\end{eqnarray}
where the former and the latter would correspond   to the section condition and the (doubled) Klein--Gordon equation respectively, while   $\hat{p}_{A}\cH^{AB}(\hat{x})\hat{p}_{B}$ is  `ordered' such that it becomes  a Hermitian operator. However,  (\ref{latter})  is not invariant under  target spacetime DFT-diffeomorphisms~\cite{Jeon:2010rw}
: the correct one should contain the DFT-dilaton, $d$
 and  read
\be
\left[-\hbar^{2}\cH^{AB}\na_{A}\na_{B}+(m\,l_{s})^{2}\right]\Psi(\hat{x})|0\rangle=
\left[-\hbar^{2}\left\{\partial_{A}\!\left(\cH^{AB}\partial_{B}\Psi\right)
-2\cH^{AB}\partial_{A}d\partial_{B}\Psi\right\}
+(m\,l_{s})^{2}\Psi\right]|0\rangle=0\,,
\label{EOMcov}
\ee
where the covariant derivative, $\na_{A}$, was defined in \cite{Jeon:2011cn} and gives the expression  after the first equality.  
{This new dilaton contribution might be comparable with  the   Fradkin--Tseytlin  term on string worldsheet,~\textit{c.f.~}\cite{Fernandez-Melgarejo:2018wpg}.
The modification \eqref{EOMcov}}  amounts to   quantum corrections  to the Hamiltonian constraint,
\be
H_{\hbar}:=\half p_{A}\cH^{AB}p_{B}+\half(m\, l_{s})^{2}+i\hbar \cH^{AB}\partial_{A}d\, p_{B}\,,
\label{hbar H}
\ee
and  accordingly to the classical  action,  
\be\dis{
S_{\hbar}=\frac{1}{\,\ls}
\int\rd\tau~\half e^{-1}\!\left(\rD_{\tau}x^{A}- i\hbar e\cH^{AC}\partial_{C}d\right)\!
\left(\rD_{\tau}x^{B}- i\hbar e\cH^{BD}\partial_{D}d\right)\!\cH_{AB}
-\half (m\,\ls)^{2}e\,.}
\label{Lhbaraction}
\ee
{It is worth while to note that  (\ref{EOMcov}) corresponds to the Euler--Lagrange equation of the following  action of the scalar field $\Psi$ which is $\ODD$ symmetric and  diffeomorphism invarant,
\be
\dis{
S[\Psi]=\int e^{-2d}\left[\cH^{AB}\partial_{A}\Psi\partial_{B}\Psi+(ml_{s}/\hbar)^{2}\Psi^{2}\right]\,.
}
\label{SPsi}
\ee
The (non-constant) dilaton appears in the equation \eqref{EOMcov} as a dissipative term.
This, in turn, is the reason for the imaginary part in the Hamiltonian \eqref{hbar H}. Nevertheless,
the scalar field action~(\ref{SPsi}) is real while the quantum particle action~(\ref{Lhbaraction}) is complex-valued.}

\section{Conclusion\label{SECconclusion}}
In this note, we have shown that the BRST formulation of the doubled-yet-gauged  particle action~(\ref{BRSTaction}) naturally produces the graded  Poisson geometry  of  \cite{Deser:2018oyg,Deser:2016qkw}.  The Grassmann odd variable, $\theta^{A}$, of the graded Poisson bracket carries an $\ODD$ vector indices. Thus, if it is to be identified as a ghost of any BRST system, the underlying gauge symmetry should be about the doubled spacetime itself, which we have shown to be the `coordinate gauge symmetry',   
$x^{A}\,\sim\,x^{A}+\cP^{A}{}_{B}V^{B}$  (\ref{EQUIV2}).    One message our work may  convey is   that, the investigation of ``spaces"  can be performed  by studying not only the functions defined on them but also the coordinate systems adopted for them, such as the doubled-yet-gauged coordinate system.

A few  comments are in order.  
With $B^{A}=\brcP^{A}{}_{B}\theta^{B}$~(\ref{BCtheta}),   if we set, instead of (\ref{Txtheta}), 
\be
\cT(x,B):=\textstyle{\frac{1}{p!}}\,T_{A_{1}A_{2}\cdots A_{p}}(x)B^{A_{1}}B^{A_{2}}\cdots B^{A_{p}}\,,
\ee
we may  realize  an  exterior derivative precisely,  \textit{c.f.~}(\ref{almost}),
\be
\left[\,p_{A}B^{A}\,,\,\cT(x,B)\right\}=-
\textstyle{\frac{1}{p!}}\,\partial_{[A_{1}}T_{A_{2}\cdots B_{p+1}]}B^{A_{1}}B^{A_{2}}\cdots B^{A_{p+1}}\,.
\label{EXTERIOR}
\ee
Now, allowing  $\ODD$ duality rotations as well as  coordinate gauge symmetry in addition to the  DFT-diffeomorphisms  for the gluing of overlapping patches, \textit{c.f.}~\cite{Kachru:2002sk,Hull:2004in}, the notion of de Rham  cohomology should differ  from the usual undoubled one.

It would be of interest for future work to generalize  { our analysis to sigma-models where}  the constant projector~(\ref{cPdef}) {is promoted} to a local object,  {\textit{c.f.~}\cite{Cederwall:2014kxa, Freidel:2017yuv,Marotta:201910}. 
Another interesting direction to pursue would be to promote} the Abelian coordinate gauge symmetry to  a non-Abelian version, as well as  to extend   this research  to  the doubled-yet-gauged string actions~\cite{Hull:2006va,Lee:2013hma,Park:2016sbw,Arvanitakis:2017hwb,Arvanitakis:2018hfn}. 

\subsection*{Acknowledgments}
We wish to thank Stephen Angus, Kyoungho Cho, and  Kevin Morand for  helpful  discussions. JHP  is also grateful to      David Berman, Ctirad Klim\v{c}\'{i}k, and Franco Pezzella for  stimulating discussions   on  doubled-yet-gauged sigma models.   This work has been 
  supported by  the National Research Foundation of Korea   through  the Grants,  
2014R1A6A3A04056670,
2016R1D1A1B01015196, 2018H1D3A1A02074698 (Korea Research Fellowship Program),  and 2019R1F1A1044065.

%
%

\bibliographystyle{utphys}
\bibliography{biblio}

\begin{thebibliography}{99}



\bibitem{Siegel:1993xq}
  W.~Siegel,
  ``Two vierbein formalism for string inspired axionic gravity,''
  Phys.\ Rev.\ D {\bf 47} (1993) 5453
  doi:10.1103/PhysRevD.47.5453
  [hep-th/9302036].



\bibitem{Siegel:1993th}
  W.~Siegel,
  ``Superspace duality in low-energy superstrings,''
  Phys.\ Rev.\ D {\bf 48} (1993) 2826
  doi:10.1103/PhysRevD.48.2826
  [hep-th/9305073].



\bibitem{Hull:2009mi}
  C.~Hull and B.~Zwiebach,
  ``Double Field Theory,''
  JHEP {\bf 0909} (2009) 099
  doi:10.1088/1126-6708/2009/09/099
  [arXiv:0904.4664 [hep-th]].



\bibitem{Hull:2009zb}
  C.~Hull and B.~Zwiebach,
  ``The Gauge algebra of double field theory and Courant brackets,''
  JHEP {\bf 0909} (2009) 090
  doi:10.1088/1126-6708/2009/09/090
  [arXiv:0908.1792 [hep-th]].

\bibitem{Hohm:2010jy}
  O.~Hohm, C.~Hull and B.~Zwiebach,
  ``Background independent action for double field theory,''
  JHEP {\bf 1007} (2010) 016
  doi:10.1007/JHEP07(2010)016
  [arXiv:1003.5027 [hep-th]].

\bibitem{Hohm:2010pp}
  O.~Hohm, C.~Hull and B.~Zwiebach,
  ``Generalized metric formulation of double field theory,''
  JHEP {\bf 1008} (2010) 008
  doi:10.1007/JHEP08(2010)008
  [arXiv:1006.4823 [hep-th]].



\bibitem{Jeon:2010rw}
  I.~Jeon, K.~Lee and J.~H.~Park,
  ``Differential geometry with a projection: Application to double field theory,''
  JHEP {\bf 1104} (2011) 014
  doi:10.1007/JHEP04(2011)014
  [arXiv:1011.1324 [hep-th]].




\bibitem{Jeon:2011cn}
  I.~Jeon, K.~Lee and J.~H.~Park,
  ``Stringy differential geometry, beyond Riemann,''
  Phys.\ Rev.\ D {\bf 84} (2011) 044022
  doi:10.1103/PhysRevD.84.044022
  [arXiv:1105.6294 [hep-th]].



  
\bibitem{Jeon:2011vx}
  I.~Jeon, K.~Lee and J.~H.~Park,
  ``Incorporation of fermions into double field theory,''
  JHEP {\bf 1111} (2011) 025
  doi:10.1007/JHEP11(2011)025
  [arXiv:1109.2035 [hep-th]].


\bibitem{Hohm:2011si}
  O.~Hohm and B.~Zwiebach,
  ``On the Riemann Tensor in Double Field Theory,''
  JHEP {\bf 1205} (2012) 126
  doi:10.1007/JHEP05(2012)126
  [arXiv:1112.5296 [hep-th]].


\bibitem{Park:2015bza}
  J.~H.~Park, S.~J.~Rey, W.~Rim and Y.~Sakatani,
  ``$\ODD$ covariant Noether currents and global charges in double field theory,''
  JHEP {\bf 1511} (2015) 131
  [arXiv:1507.07545 [hep-th]].




\bibitem{Rocen:2010bk}
  A.~Rocen and P.~West,
  ``E11, generalised space-time and IIA string theory: the R-R sector,''
  doi:10.1142/9789814412551$\underline{\,\,\,}$0020
  arXiv:1012.2744 [hep-th].
  
  




\bibitem{Hohm:2011ex}
  O.~Hohm and S.~K.~Kwak,
  ``Double Field Theory Formulation of Heterotic Strings,''
  JHEP {\bf 1106} (2011) 096
  doi:10.1007/JHEP06(2011)096
  [arXiv:1103.2136 [hep-th]].


  
\bibitem{Hohm:2011zr}
  O.~Hohm, S.~K.~Kwak and B.~Zwiebach,
  ``Unification of Type II Strings and T-duality,''
  Phys.\ Rev.\ Lett.\  {\bf 107} (2011) 171603
  doi:10.1103/PhysRevLett.107.171603
  [arXiv:1106.5452 [hep-th]].
    
\bibitem{Hohm:2011dv}
  O.~Hohm, S.~K.~Kwak and B.~Zwiebach,
  ``Double Field Theory of Type II Strings,''
  JHEP {\bf 1109} (2011) 013
  doi:10.1007/JHEP09(2011)013
  [arXiv:1107.0008 [hep-th]].


  
  

\bibitem{Jeon:2012kd}
  I.~Jeon, K.~Lee and J.~H.~Park,
  ``Ramond-Ramond Cohomology and O(D,D) T-duality,''
  JHEP {\bf 1209} (2012) 079
  doi:10.1007/JHEP09(2012)079
  [arXiv:1206.3478 [hep-th]].
  
  

 




\bibitem{Jeon:2011kp}
  I.~Jeon, K.~Lee and J.~H.~Park,
  ``Double field formulation of Yang--Mills theory,''
  Phys.\ Lett.\ B {\bf 701} (2011) 260
  doi:10.1016/j.physletb.2011.05.051
  [arXiv:1102.0419 [hep-th]].

\bibitem{Choi:2015bga}
  K.~S.~Choi and J.~H.~Park,
  ``Standard Model as a Double Field Theory,''
  Phys.\ Rev.\ Lett.\  {\bf 115} (2015) no.17,  171603
  doi:10.1103/PhysRevLett.115.171603
  [arXiv:1506.05277 [hep-th]].



\bibitem{Bekaert:2016isw}
  X.~Bekaert and J.~H.~Park,
  ``Higher Spin Double Field Theory : A Proposal,''
  JHEP {\bf 1607} (2016) 062
  doi:10.1007/JHEP07(2016)062
  [arXiv:1605.00403 [hep-th]].


\bibitem{Angus:2018mep}
  S.~Angus, K.~Cho and J.~H.~Park,
  ``Einstein Double Field Equations,''
  Eur.\ Phys.\ J.\ C {\bf 78} (2018) no.6,  500
  doi:10.1140/epjc/s10052-018-5982-y
  [arXiv:1804.00964 [hep-th]].





\bibitem{Park:2019hbc}
  J.~H.~Park,
  ``$\mathbf{O}(D,D)$ completion of the Einstein Field Equations,''
  arXiv:1904.04705 [hep-th].


\bibitem{Jeon:2011sq}
  I.~Jeon, K.~Lee and J.~H.~Park,
  ``Supersymmetric Double Field Theory: Stringy Reformulation of Supergravity,''
  Phys.\ Rev.\ D {\bf 85} (2012) 081501
  [arXiv:1112.0069 [hep-th]].
  


\bibitem{Jeon:2012hp}
  I.~Jeon, K.~Lee, J.~H.~Park and Y.~Suh,
  ``Stringy Unification of Type IIA and IIB Supergravities under N=2 D=10 Supersymmetric Double Field Theory,''
  Phys.\ Lett.\ B {\bf 723} (2013) 245
  doi:10.1016/j.physletb.2013.05.016
  [arXiv:1210.5078 [hep-th]].
  
  
  

\bibitem{Morand:2017fnv}
  K.~Morand and J.~H.~Park,
  ``Classification of non-Riemannian doubled-yet-gauged spacetime,''
  Eur.\ Phys.\ J.\ C {\bf 77} (2017) no.10,  685
  [arXiv:1707.03713 [hep-th]].



\bibitem{Berman:2019izh}
  D.~S.~Berman, C.~D.~A.~Blair and R.~Otsuki,
  ``Non-Riemannian geometry of M-theory,''
  JHEP {\bf 1907} (2019) 175
  doi:10.1007/JHEP07(2019)175
  [arXiv:1902.01867 [hep-th]].


\bibitem{Blair:2019qwi}
  C.~D.~A.~Blair,
  ``A worldsheet supersymmetric Newton-Cartan string,''
  arXiv:1908.00074 [hep-th].

\bibitem{Cho:2019ofr}
  K.~Cho and J.~H.~Park,
  ``Remarks on the non-Riemannian sector in Double Field Theory,''
  arXiv:1909.10711 [hep-th].




\bibitem{Park:2013mpa}
  J.~H.~Park,
  ``Comments on double field theory and diffeomorphisms,''
  JHEP {\bf 1306} (2013) 098
  doi:10.1007/JHEP06(2013)098
  [arXiv:1304.5946 [hep-th]].

  
\bibitem{Hull:2006va}
  C.~M.~Hull,
  ``Doubled Geometry and T-Folds,''
  JHEP {\bf 0707} (2007) 080
  [hep-th/0605149].


\bibitem{Hull:2009sg}
  C.~M.~Hull and R.~A.~Reid-Edwards,
  ``Non-geometric backgrounds, doubled geometry and generalised T-duality,''
  JHEP {\bf 0909} (2009) 014
  doi:10.1088/1126-6708/2009/09/014
  [arXiv:0902.4032 [hep-th]].



\bibitem{Lee:2013hma}
  K.~Lee and J.~H.~Park,
  ``Covariant action for a string in doubled yet gauged  spacetime,''
  Nucl.\ Phys.\ B {\bf 880} (2014) 134 
  [arXiv:1307.8377 [hep-th]].
  


\bibitem{Bakas:2016nxt}
  I.~Bakas, D.~L\"{u}st and E.~Plauschinn,
  ``Towards a world‐sheet description of doubled geometry in string theory,''
  Fortsch.\ Phys.\  {\bf 64} (2016) no.10,  730
  [arXiv:1602.07705 [hep-th]].




\bibitem{Ko:2016dxa}
  S.~M.~Ko, J.~H.~Park and M.~Suh,
  ``The rotation curve of a point particle in stringy gravity,''
  JCAP {\bf 1706} (2017) 002
  doi:10.1088/1475-7516/2017/06/002
  [arXiv:1606.09307 [hep-th]].


\bibitem{Park:2016sbw}
  J.~H.~Park,
  ``Green-Schwarz superstring on doubled-yet-gauged spacetime,''
  JHEP {\bf 1611} (2016) 005
  [arXiv:1609.04265 [hep-th]].

\bibitem{Arvanitakis:2017hwb}
  A.~S.~Arvanitakis and C.~D.~A.~Blair,
  ``Unifying Type-II Strings by Exceptional Groups,''
  Phys.\ Rev.\ Lett.\  {\bf 120} (2018) no.21,  211601
  [arXiv:1712.07115 [hep-th]].


\bibitem{Arvanitakis:2018hfn}
  A.~S.~Arvanitakis and C.~D.~A.~Blair,
  ``The Exceptional Sigma Model,''
  JHEP {\bf 1804} (2018) 064
  [arXiv:1802.00442 [hep-th]].



\bibitem{Marotta:2018swj}
  V.~E.~Marotta, F.~Pezzella and P.~Vitale,
  ``Doubling, T-Duality and Generalized Geometry: a Simple Model,''
  JHEP {\bf 1808} (2018) 185
  doi:10.1007/JHEP08(2018)185
  [arXiv:1804.00744 [hep-th]].
  
\bibitem{Marotta:2019wfq}
  V.~E.~Marotta, F.~Pezzella and P.~Vitale,
  ``T-Dualities and Doubled Geometry of Principal Chiral Model,''
  arXiv:1903.01243 [hep-th].

\bibitem{Bascone:2019tuc}
  F.~Bascone, V.~E.~Marotta, F.~Pezzella and P.~Vitale,
  ``T-Duality and Doubling of the Isotropic Rigid Rotator,''
  PoS CORFU {\bf 2018} (2019) 123
  doi:10.22323/1.347.0123
  [arXiv:1904.03727 [hep-th]].





\bibitem{Hitchin:2004ut}
  N.~Hitchin,
  ``Generalized Calabi-Yau manifolds,''
  Quart.\ J.\ Math.\  {\bf 54} (2003) 281
  doi:10.1093/qjmath/54.3.281
  [math/0209099 [math-dg]].



\bibitem{Gualtieri:2003dx}
  M.~Gualtieri,
  ``Generalized complex geometry,''
  math/0401221 [math-dg].



\bibitem{Hitchin:2010qz}
  N.~Hitchin,
  ``Lectures on generalized geometry,''
  arXiv:1008.0973 [math.DG].




\bibitem{Cavalcanti:2011wu}
 G. R. Cavalcanti and M. Gualtieri,
  ``Generalized complex geometry and T-duality,''
  A Celebration of the Mathematical Legacy of Raoul Bott (CRM
  Proceedings \& Lecture Notes) American Mathematical Society (2010)
  [arXiv:1106.1747].


\bibitem{Coimbra:2011nw}
  A.~Coimbra, C.~Strickland-Constable and D.~Waldram,
  ``Supergravity as Generalised Geometry I: Type II Theories,''
  JHEP {\bf 1111} (2011) 091
  doi:10.1007/JHEP11(2011)091
  [arXiv:1107.1733 [hep-th]].



\bibitem{Coimbra:2012yy}
  A.~Coimbra, C.~Strickland-Constable and D.~Waldram,
  ``Generalised Geometry and type II Supergravity,''
  Fortsch.\ Phys.\  {\bf 60} (2012) 982
  doi:10.1002/prop.201100096
  [arXiv:1202.3170 [hep-th]].




\bibitem{Chatzistavrakidis:2019huz}
  A.~Chatzistavrakidis, L.~Jonke, F.~S.~Khoo and R.~J.~Szabo,
  ``The Algebroid Structure of Double Field Theory,''
  PoS CORFU {\bf 2018} (2019) 132
  doi:10.22323/1.347.0132
  [arXiv:1903.01765 [hep-th]].

\bibitem{Chatzistavrakidis:2018ztm}
  A.~Chatzistavrakidis, L.~Jonke, F.~S.~Khoo and R.~J.~Szabo,
  ``Double Field Theory and Membrane Sigma-Models,''
  JHEP {\bf 1807} (2018) 015
  doi:10.1007/JHEP07(2018)015
  [arXiv:1802.07003 [hep-th]].



\bibitem{Chatzistavrakidis:2019rpp}
  A.~Chatzistavrakidis, C.~J.~Grewcoe, L.~Jonke, F.~S.~Khoo and R.~J.~Szabo,
  ``BRST symmetry of doubled membrane sigma-models,''
  PoS CORFU {\bf 2018} (2019) 147
  [arXiv:1904.04857 [hep-th]].



\bibitem{Hull:2004in}
  C.~M.~Hull,
  ``A Geometry for non-geometric string backgrounds,''
  JHEP {\bf 0510} (2005) 065
  doi:10.1088/1126-6708/2005/10/065
  [hep-th/0406102].



\bibitem{Vaisman:2012ke}
  I.~Vaisman,
  ``On the geometry of double field theory,''
  J.\ Math.\ Phys.\  {\bf 53} (2012) 033509
  doi:10.1063/1.3694739
  [arXiv:1203.0836 [math.DG]].


\bibitem{Vaisman:2012px}
  I.~Vaisman,
  ``Towards a double field theory on para-Hermitian manifolds,''
  J.\ Math.\ Phys.\  {\bf 54} (2013) 123507
  doi:10.1063/1.4848777
  [arXiv:1209.0152 [math.DG]].







\bibitem{Freidel:2017yuv}
  L.~Freidel, F.~J.~Rudolph and D.~Svoboda,
  ``Generalised Kinematics for Double Field Theory,''
  JHEP {\bf 1711} (2017) 175
  doi:10.1007/JHEP11(2017)175
  [arXiv:1706.07089 [hep-th]].

\bibitem{Svoboda:2018rci}
  D.~Svoboda,
  ``Algebroid Structures on Para-Hermitian Manifolds,''
  J.\ Math.\ Phys.\  {\bf 59} (2018) no.12,  122302
  doi:10.1063/1.5040263
  [arXiv:1802.08180 [math.DG]].

\bibitem{Freidel:2018tkj}
  L.~Freidel, F.~J.~Rudolph and D.~Svoboda,
  ``A Unique Connection for Born Geometry,''
  doi:10.1007/s00220-019-03379-7
  arXiv:1806.05992 [hep-th].



\bibitem{Marotta:2018myj}
  V.~E.~Marotta and R.~J.~Szabo,
  ``Para‐ Hermitian Geometry, Dualities and Generalized Flux Backgrounds,''
  Fortsch.\ Phys.\  {\bf 67} (2019) no.3,  1800093
  [arXiv:1810.03953 [hep-th]].

\bibitem{Mori:2019slw}
  H.~Mori, S.~Sasaki and K.~Shiozawa,
  ``Doubled Aspects of Vaisman Algebroid and Gauge Symmetry in Double Field Theory,''
  arXiv:1901.04777 [hep-th].



\bibitem{Hassler:2019wvn}
  F.~Hassler, D.~L\"{u}st and F.~J.~Rudolph,
  ``Para-Hermitian Geometries for Poisson-Lie Symmetric $\sigma$-models,''
  arXiv:1905.03791 [hep-th].



\bibitem{Marotta:201910}
V.~E.~Marotta and R.~J.~Szabo,
``Born Sigma-Models for Para-Hermitian Manifolds and Generalized T-Duality,"
 arXiv:1910.09997 [hep-th].



\bibitem{Heller:2016abk}
  M.~A.~Heller, N.~Ikeda and S.~Watamura,
  ``Unified picture of non-geometric fluxes and T-duality in double field theory via graded symplectic manifolds,''
  JHEP {\bf 1702} (2017) 078
  [arXiv:1611.08346].


\bibitem{Heller:2017mwz}
  M.~A.~Heller, N.~Ikeda and S.~Watamura,
  ``Courant algebroids from double field theory in supergeometry,''
  doi:10.1142/9789813144613$\underline{\,\,\,}$0008
  arXiv:1703.00638 [hep-th].







  



\bibitem{Deser:2014mxa}
  A.~Deser and J.~Stasheff,
  ``Even symplectic supermanifolds and double field theory,''
  Commun.\ Math.\ Phys.\  {\bf 339} (2015) no.3,  1003
  doi:10.1007/s00220-015-2443-4
  [arXiv:1406.3601 [math-ph]].




\bibitem{Deser:2016qkw}
  A.~Deser and C.~S\"{a}mann,
  ``Extended Riemannian Geometry I: Local Double Field Theory,''
  doi:10.1007/s00023-018-0694-2
  arXiv:1611.02772 [hep-th].
  
 
\bibitem{Deser:2018oyg}
  A.~Deser and C.~S\"{a}mann,
  ``Derived Brackets and Symmetries in Generalized Geometry and Double Field Theory,''
  PoS CORFU {\bf 2017} (2018) 141
  doi:10.22323/1.318.0141
  [arXiv:1803.01659 [hep-th]].
    
  
\bibitem{Deser:2017fko}
  A.~Deser, M.~A.~Heller and C.~S\"{a}mann,
  ``Extended Riemannian Geometry II: Local Heterotic Double Field Theory,''
  JHEP {\bf 1804} (2018) 106
  doi:10.1007/JHEP04(2018)106
  [arXiv:1711.03308 [hep-th]].
  
  

  

\bibitem{Stasheff:2018vnl}
  J.~Stasheff,
  ``$L_\infty$ and $A_\infty$ structures: then and now,''
  arXiv:1809.02526 [math.QA].

  
\bibitem{Deser:2018flj}
  A.~Deser and C.~S\"{a}mann,
  ``Extended Riemannian Geometry III: Global Double Field Theory with Nilmanifolds,''
  JHEP {\bf 1905} (2019) 209
  doi:10.1007/JHEP05(2019)209
  [arXiv:1812.00026 [hep-th]].



\bibitem{KosmannSchwarzbach1996}
Y. Kosmann-Schwarzbach, ``From poisson algebras to gerstenhaber algebras,''  Ann. Inst. Fourier 46 (1996) 1243–1274.



\bibitem{KosmannSchwarzbach:2003en}
  Y. Kosmann-Schwarzbach,
  ``Derived brackets,''
  Lett.\ Math.\ Phys.\  {\bf 69} (2004) 61
  doi:10.1007/s11005-004-0608-8
  [arXiv:math/0312524 [math-dg]].





\bibitem{Voronov2005}
T.  Voronov, 
  ``Higher derived brackets and homotopy algebras,''
  Journal of Pure and Applied Algebra
  doi:10.1016/j.jpaa.2005.01.010
  [arXiv:math/0304038].


\bibitem{Voronov2004}
 T. Voronov,  
  ``Higher derived brackets for arbitrary derivations,''
  Travaux Math\'ematiques XVI
  [arXiv:math/041220].



\bibitem{Getzler2010}
E. Getzler,   ``Higher derived brackets,''
  [arXiv:1010.5859].
  
\bibitem{Alfonsi:2019ggg}
  L.~Alfonsi,
  ``Global Double Field Theory is Higher Kaluza-Klein Theory,''
  arXiv:1912.07089 [hep-th].


\bibitem{Hohm:2012gk}
  O.~Hohm and B.~Zwiebach,
  ``Large Gauge Transformations in Double Field Theory,''
  JHEP {\bf 1302} (2013) 075
  doi:10.1007/JHEP02(2013)075
  [arXiv:1207.4198 [hep-th]].

\bibitem{Berman:2014jba}
  D.~S.~Berman, M.~Cederwall and M.~J.~Perry,
  ``Global aspects of double geometry,''
  JHEP {\bf 1409} (2014) 066
  doi:10.1007/JHEP09(2014)066
  [arXiv:1401.1311 [hep-th]].



\bibitem{Hull:2014mxa}
  C.~M.~Hull,
  ``Finite Gauge Transformations and Geometry in Double Field Theory,''
  JHEP {\bf 1504} (2015) 109
  doi:10.1007/JHEP04(2015)109
  [arXiv:1406.7794 [hep-th]].




\bibitem{Cederwall:2014opa}
  M.~Cederwall,
   ``T-duality and non-geometric solutions from double geometry,''
  Fortsch.\ Phys.\  {\bf 62} (2014) 942
  doi:10.1002/prop.201400069
  [arXiv:1409.4463 [hep-th]].






\bibitem{Park:2017snt}
  J.~H.~Park,
  ``Stringy Gravity: Solving the Dark Problems at ‘short’ distance,''
  EPJ Web Conf.\  {\bf 168} (2018) 01010
  doi:10.1051/epjconf/201816801010
  [arXiv:1707.08961 [hep-th]].



  





\bibitem{Henneaux:1992ig}
  M.~Henneaux, C.~Teitelboim,
  ``Quantization of gauge systems,''
  Princeton, USA: Univ. Pr. (1992) 520p.




\bibitem{Bergshoeff:2019sfy}
  E.~Bergshoeff, A.~Kleinschmidt, E.~T.~Musaev and F.~Riccioni,
  ``The different faces of branes in Double Field Theory,''
  JHEP {\bf 1909} (2019) 110
  [arXiv:1903.05601 [hep-th]].



\bibitem{Fernandez-Melgarejo:2018wpg}
  J.~J.~Fernández-Melgarejo, J.~I.~Sakamoto, Y.~Sakatani and K.~Yoshida,
  ``Weyl invariance of string theories in generalized supergravity backgrounds,''
  Phys.\ Rev.\ Lett.\  {\bf 122} (2019) no.11,  111602
  [arXiv:1811.10600 [hep-th]].


\bibitem{Kachru:2002sk}
  S.~Kachru, M.~B.~Schulz, P.~K.~Tripathy and S.~P.~Trivedi,
  ``New supersymmetric string compactifications,''
  JHEP {\bf 0303} (2003) 061
  doi:10.1088/1126-6708/2003/03/061
  [hep-th/0211182].






\bibitem{Cederwall:2014kxa}
  M.~Cederwall,
  ``The geometry behind double geometry,''
  JHEP {\bf 1409} (2014) 070
  doi:10.1007/JHEP09(2014)070
  [arXiv:1402.2513 [hep-th]].






\end{thebibliography}

\end{document}